\documentclass[10pt,twocolumn,twoside,letterpaper]{IEEEtran}

\usepackage{geometry}
\usepackage{xcolor}
\usepackage{graphicx} 
\graphicspath{{images/}} 
\usepackage{amsmath} 
\usepackage{amssymb} 
\usepackage{adjustbox}
\usepackage{booktabs} 
\usepackage{enumitem} 
\usepackage{palatino} 
\usepackage[font=small,labelfont=bf]{caption} 
\usepackage{url}
\usepackage{multicol} 
\usepackage{wrapfig}
\usepackage{subcaption}
\usepackage[absolute]{textpos}
\makeatletter
\def\ps@IEEEtitlepagestyle{
  \def\@oddfoot{\mycopyrightnotice}
  \def\@evenfoot{}
}
\def\mycopyrightnotice{
  {\footnotesize
  \begin{minipage}{\textwidth}
  \centering
  978-1-7281-7693-2/20/\$31.00 \copyright2020 IEEE
  \end{minipage}
  }
}

\usepackage{url}

\begin{document}

\title{Investigation and Mitigation of Crosstalk in the Prototype ME0 GEM Detector for the Phase-2 Muon System Upgrade of the CMS Experiment}
%
%
%

\author{Stephen~D.~Butalla and Marcus Hohlmann,\\\emph{On behalf of the CMS Muon Group}
\thanks{Manuscript received December 20, 2020. This work was supported in part by the U.S. Department of Energy Office of Science (HEP) and the National Science Foundation (Cornell University Subaward; P.I. Dr. Marcus Hohlmann).}
\thanks{S. D. Butalla and M. Hohlmann are with the Aerospace, Physics and Space Sciences Department at Florida Institute of Technology, Melbourne FL, 32901 (email: sbutalla2012@my.fit.edu and hohlmann@fit.edu).}}

\maketitle

\pagenumbering{gobble}
\begin{textblock}{18}(2,1)\center \vspace{-0.4cm}
\noindent This work has been submitted to the conference proceedings of the 2020 IEEE NSS-MIC Conference for publication. Copyright may be transferred without notice, after which this version may no longer be available.
\end{textblock}

\begin{abstract}
The LHC is currently undergoing a high luminosity upgrade, which is set to increase the instantaneous luminosity by at least a factor of five. This luminosity increase will result in a higher muon flux rate in the forward region and overwhelm the current trigger system of the CMS experiment. The ME0, a gas electron multiplier detector, is proposed for the Phase-2 Muon System Upgrade for the CMS experiment to help increase the muon acceptance and to control the Level 1 muon trigger rate. A recent design iteration of this detector features GEM foils that are segmented on both sides, which helps to lower the probability of high voltage discharges. However, during preliminary testing of the chamber, substantial crosstalk between readout sectors was observed. Here, we investigate, characterize, and quantify the crosstalk present in the detector, and also estimate the performance of the chamber as a result of this crosstalk via simulation results of the detector dead time, efficiency loss, and frontend electronics response. The results of crosstalk via signals produced by applying a square voltage pulse directly on the readout strips of the detector with a signal generator are summarized. We also present the efficacy of mitigation strategies including bypass capacitors and increasing the area of the HV segments on the third GEM foil in the detector. We find that the crosstalk is a result of capacitive coupling between the readout strips on the readout board and between the readout strips and the bottom of the third GEM foil. Our results show that the crosstalk generally follows a pattern where the largest magnitude of crosstalk is within the same azimuthal readout segment in the detector, and in the next-nearest horizontal segments in eta. Generally, the bypass capacitors and increased area of the HV segments successfully lower the crosstalk in the sectors where they are located; on average, we observe a maximum decrease of crosstalk in sectors previously experiencing crosstalk from $(1.66\pm0.03)\%$ to $(1.11\pm0.02)\%$ with all HV segments connected in parallel on the bottom of the third GEM foil, with the addition of an HV low-pass filter connected to this electrode, and an HV divider. However, with these mitigation strategies, we also observe slightly increased crosstalk $\big(\hspace{-0.1cm}\lessapprox 0.4\%\big)$ in readout sectors farther away.
\end{abstract}

\newpage

\section{INTRODUCTION}
\begin{figure}[!ht]
\centering
\includegraphics[width=\columnwidth]{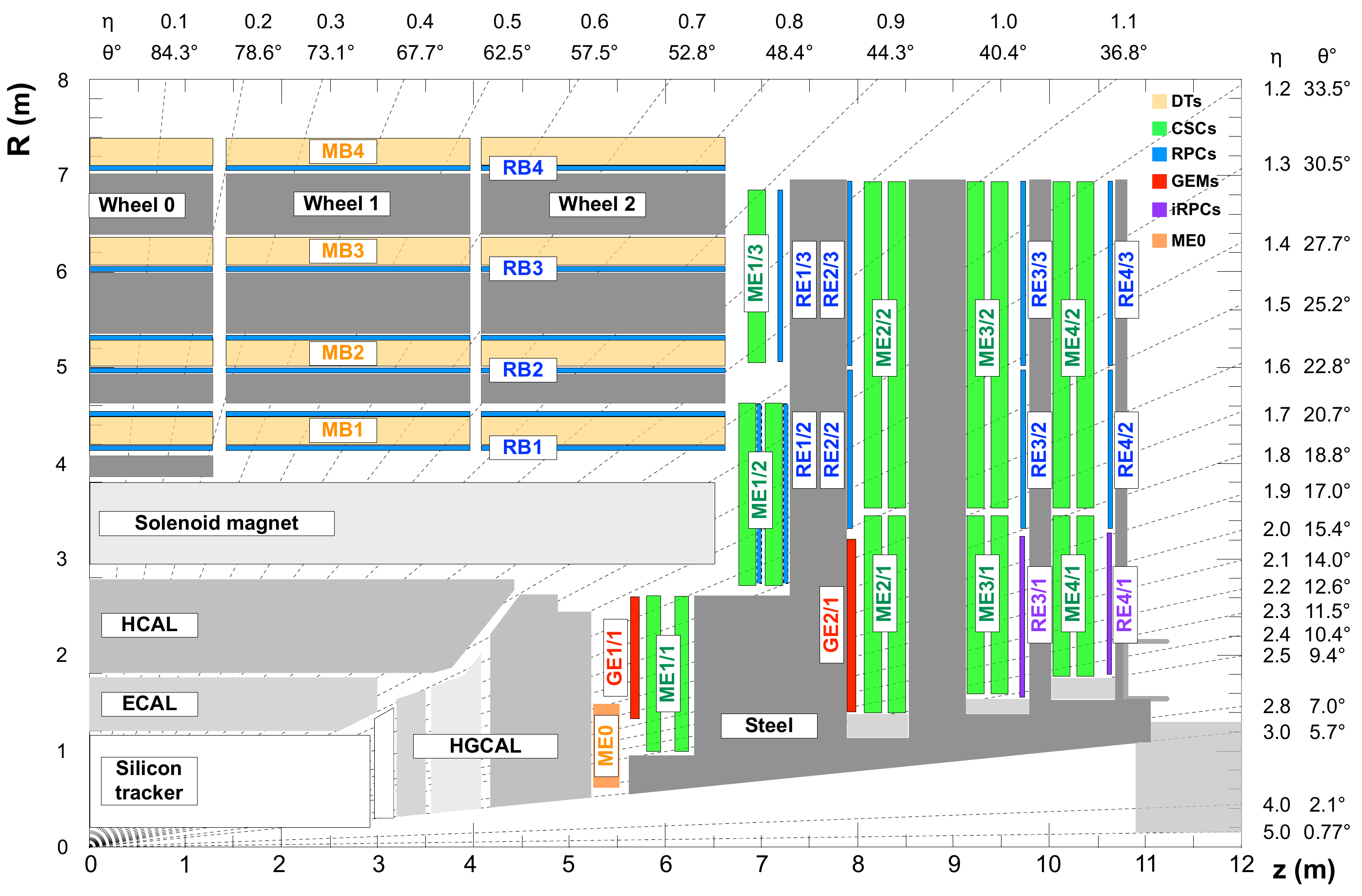}
\caption{\label{fig:quadrant}Quadrant of the upgraded CMS experiment with the ME0 in orange \cite{TDR}.}
\end{figure}

%
%
%
%
\IEEEPARstart{T}{he} high luminosity upgrade of the LHC at CERN is projected to increase the instantaneous design luminosity by at least a factor of five. In order to cope with the increased muon flux rates from this higher luminosity, the CMS experiment is undergoing the Phase-2 Muon System upgrade \cite{TDR}, which will increase the redundancy of the muon system, as well as lower the Level 1 trigger rate. One of the detectors to be installed during the upgrade is the ME0 triple-Gas Electron Multiplier (GEM) detector, which will increase the muon acceptance from $|2.4|$ to $|2.8|$ in pseudorapidity (see Fig.~\ref{fig:quadrant}), and also help control the Level 1 trigger rate.

\begin{figure}[!ht]
\centering
\includegraphics[width=\columnwidth]{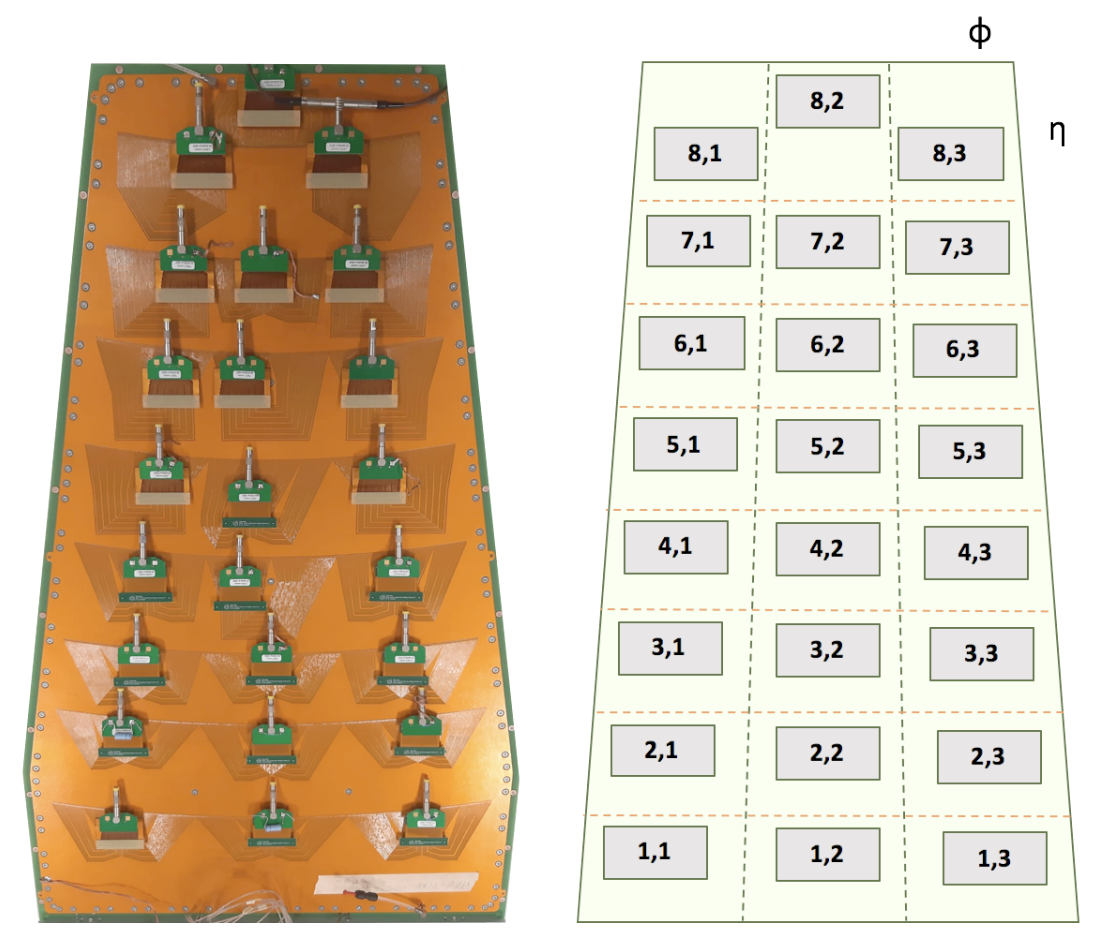}
\caption{\label{fig:ME0}The CMS ME0 triple-GEM detector and its readout sector labeling. Sector numbering ranges azimuthally from 1 to 3 and vertically from 1 to 8.}
\end{figure}

The recent design of this detector differs from a previous generation of CMS GEM detectors, the GE1/1, in that it features GEM foils with high voltage (HV) segments with protection resistors on both sides of the foils, which function to protect the chamber from HV discharges. By contrast, only the side of the GEM foils in GE1/1 detectors facing the drift electrode is segmented. The foils are divided into 37 HV segments with a range of areas from 98.7 cm\textsuperscript{2} to 103.4 cm\textsuperscript{2}. During the quality control testing of the first prototype, we observed crosstalk in neighboring readout sectors. This paper provides a summary of the investigations of the crosstalk in the ME0 detector. We characterize and quantify the crosstalk pulses in this detector by injecting signal pulses into the readout sectors with a signal generator (see Fig.~\ref{fig:ME0} for a picture of the ME0 GEM detector and the associated readout sector partitioning). We also discuss simulation results of expected detector efficiency loss, which are based on results from crosstalk rate measurements using pulses created by alpha and beta sources, background rate simulations, and dead time simulations from the response of the frontend application specific integrated circuit (ASIC) mounted on hybrid cards due to the crosstalk. Finally, we discuss mitigation techniques including the use of bypass capacitors on the GEM foil and increasing the area of the HV segments on the third GEM foil to reduce the magnitude of the crosstalk in other readout segments.

\section{Characterization of the Crosstalk}
\begin{figure}[!ht]
\centering
\includegraphics[width=\columnwidth]{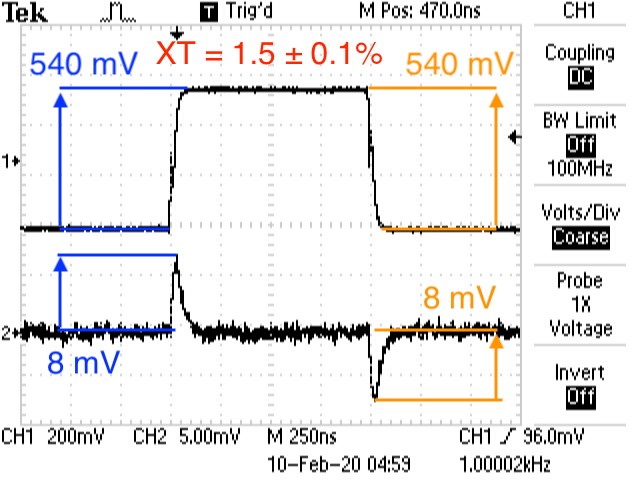}
\caption{\label{fig:traces}An example oscilloscope trace of the input square pulse (channel 1, top) and the crosstalk signal (channel 2, bottom).}
\end{figure}

During the preliminary alpha-irradiation testing of the prototype ME0 detector, we observed bipolar pulses in sectors that were not being irradiated. To investigate further, we injected a square pulse of 1 microsecond width\footnote{A typical pulse width in a GEM detector is on the order of 10 ns. However, with the capabilities of the signal generator used in this study, the amplitude of the pulse was preserved, but became distorted due to an impedance mismatch below one microsecond. Thus, we used a one microsecond pulse width for these studies.} into a readout sector (128 readout strips), and then read out the crosstalk signal on 128 ganged strips in all of the other readout sectors with an oscilloscope. To measure and quantify the crosstalk, these oscilloscope traces were recorded and their amplitudes manually measured. Figure \ref{fig:traces} displays such a measurement: on channel 1 of the trace (top), we see the injected square pulse, and on channel 2 (bottom), we see the resulting crosstalk pulse in an adjacent readout sector.

The crosstalk percentage is quantified as the ratio of output pulse amplitude to the input amplitude, multiplied by 100\%, with the error given by the standard error propagation formula below. These tests showed a maximum crosstalk amplitude of $(6.40\pm0.42)\%$ of the injected amplitude.

\begin{align}
XT\%		&= \dfrac{V_{\textrm{out}}}{V_{\textrm{in}}}\cdot100\% \label{eqn:XT}\\ 
\delta(XT\%)&=|XT|\sqrt{\bigg(\dfrac{\delta V_{\textrm{in}}}{V_{\textrm{in}}}\bigg)^{2}+\bigg(\dfrac{\delta V_{\textrm{out}}}{V_{\textrm{out}}}\bigg)^{2}}\cdot 100\%\label{eqn:XTerr}
\end{align}

\begin{figure}[!ht]
\centering
\includegraphics[width=0.8\columnwidth]{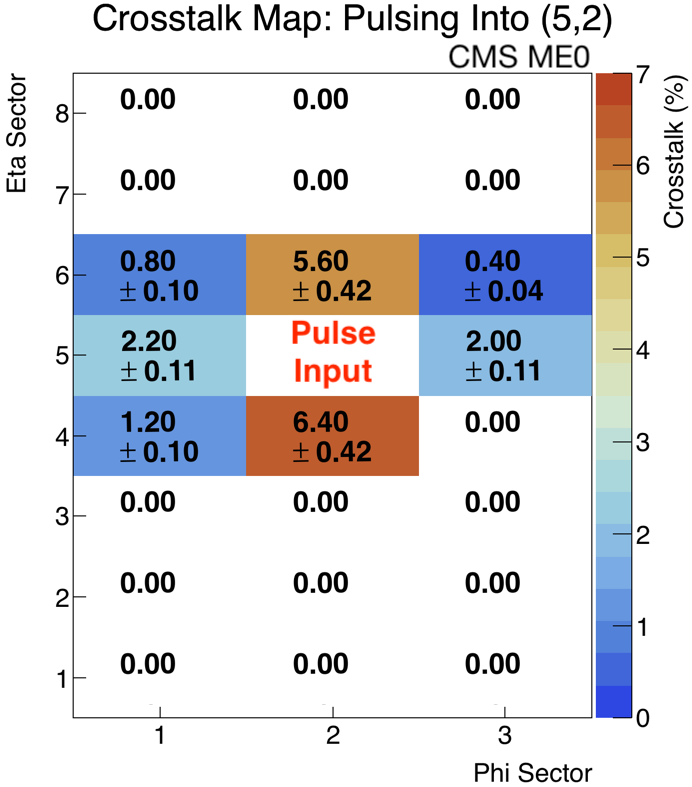}
\caption{\label{fig:map}An example crosstalk map with pulse input in sector $(\eta=5,\phi=2)$. Note the nearly symmetric behavior in adjacent $\phi$ partitions (and $\eta$ partitions). Sectors with $XT = 0.00\%$ showed no discernible crosstalk.}
\end{figure}

Comprehensive crosstalk ``maps" (see an example in Fig.~\ref{fig:map}) were made by reading out the signal in all of the other readout (RO) sectors in the chamber. These maps identify the largest observed crosstalk and the extent to which neighboring sectors in the chamber are affected. Sectors with 0.00\% crosstalk were those that did not display a crosstalk signal differentiable from baseline. Consequently, the error listed for these sectors is undefined as prescribed by \eqref{eqn:XTerr}. The range of the average observed crosstalk across the detector for pulse inputs into each $\phi$ partition of three $\eta$ segments are listed in Table \ref{tab:range}.

\begin{table}[!ht]
\renewcommand{\arraystretch}{1.3}
\caption{\label{tab:range}Range of Observed Crosstalk for Pulse Inputs Into Each of the Three Phi Partitions of the Listed Eta Segments.}
\centering
\begin{tabular}{ccc}
\hline
$\eta$ Sector & Minimum Crosstalk (\%) & Maximum Crosstalk (\%)\\
\hline
1 & 0.24$\pm$0.04 & 3.80$\pm$0.21\\
5 & 0.20$\pm$0.04 & 6.40$\pm$0.42 \\
8 & 0.16$\pm$0.04 & 4.00$\pm$0.22\\
\hline
\end{tabular}
\end{table}

\begin{figure}[!ht]
\centering
\includegraphics[width=0.72\columnwidth]{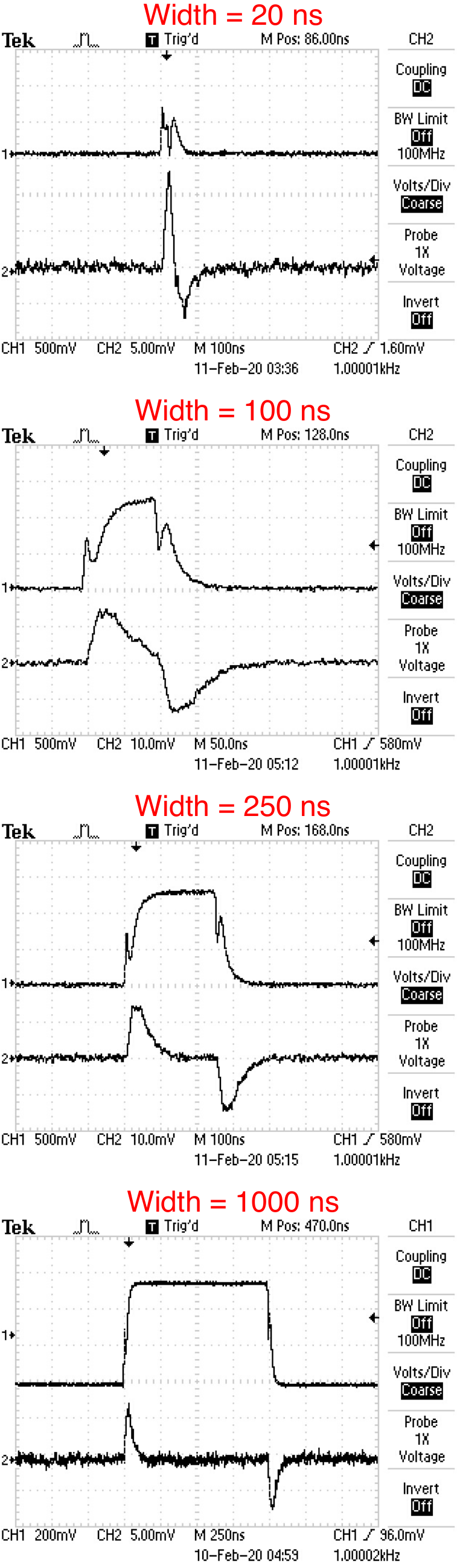}
\caption{\label{fig:comparison}Crosstalk pulse shapes for different input square pulse widths. Note that below 1 $\mu$s, the input pulses were distorted due to impedance mismatch between the cable and the strip sector.}
\end{figure}

The crosstalk signal is a result of CR differentiation: the capacitive coupling between RO sectors (and the coupling due to the bottom electrode of GEM3 that faces the strips) results in an average measured intersector capacitance $C = 702 \pm 18$ pF), and the resistance is that of the 50 $\Omega$ characteristic impedance of the cable. The time constant is then $\tau\approx$ 35 ns. This hypothesis was verified by examining the time constant of the observed crosstalk pulses and also by varying the input square pulse widths $T$, which shows the characteristic behavior of a CR differentiator for pulse widths $T \gg \tau$ and $T \approx \tau$ (see Fig.~\ref{fig:comparison}). These results are also verified via circuit simulations reported in a complementary paper by M. Hohlmann submitted to these proceedings, see \cite{Hohlmann}.

\section{Estimating the Impact of the Crosstalk on Detector Performance}
\begin{figure}[!ht]
\centering
\includegraphics[width=\columnwidth]{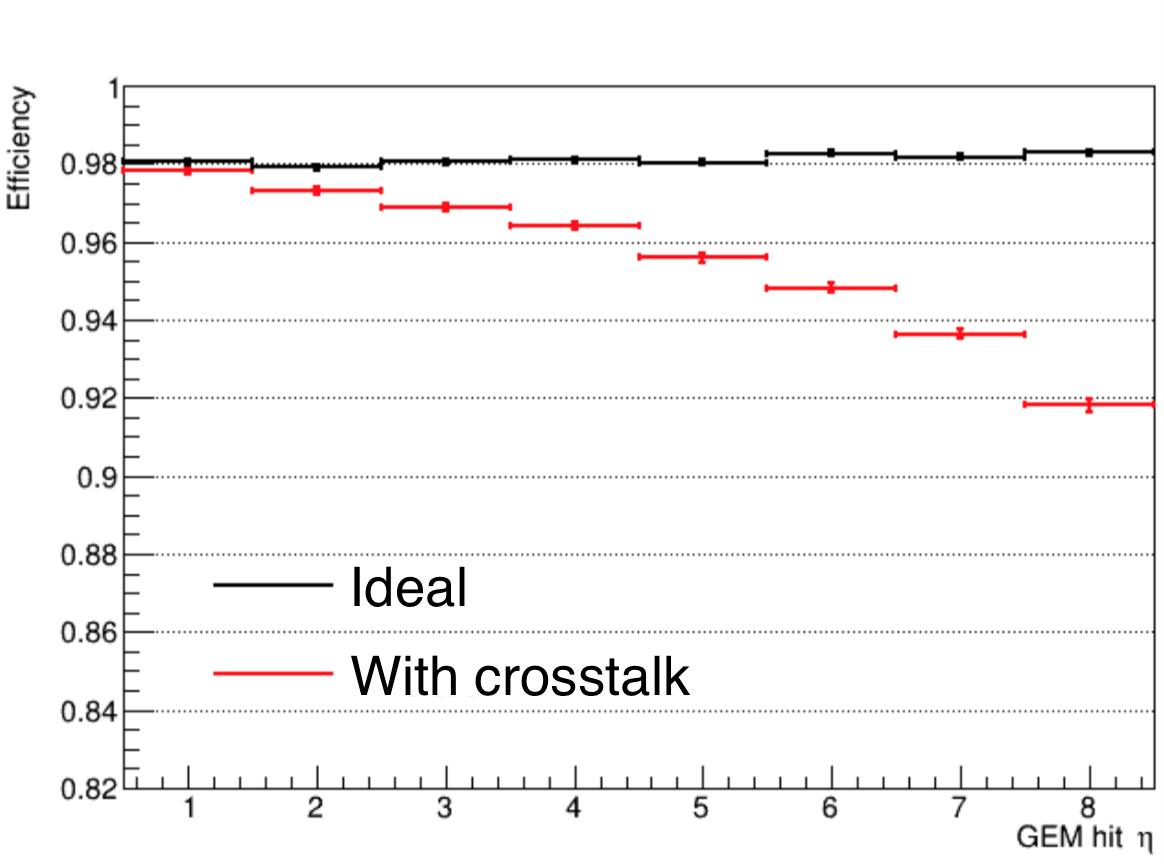}
\caption{\label{fig:simulation}Plot of the simulated reconstruction efficiencies and the losses due to crosstalk with increasing pseudorapidity (for an impact time equivalent to 50 bunch crossings).}
\end{figure}

To estimate the impact that this crosstalk has on detector performance, experimental results determining the probability of observing a crosstalk pulse were used as input parameters to a simulation of the ensuing background rate in the detector from crosstalk.. To determine the crosstalk probability, a GE1/1 GEM detector with double-segmented foils was irradiated with alpha and beta sources through a small hole in the GEM drift cathode PCB, and the hit rate of the pulses above threshold were recorded. Dead time and timing error simulations of the frontend ASIC hybrid cards were performed by injecting a signal pulse at a fixed time into the simulated shaping circuit, and then varying the injection time of a crosstalk pulse into the same, simulated circuit of the ASIC. It was found that a maximum timing error of about 550 ns results from the interference of the large crosstalk signal with actual signal pulses on neighboring strips. Results of heavily-ionizing background particle rate simulation in CMS were used in tandem with the simulations to determine the loss of efficiency of the detector. Figure \ref{fig:simulation} displays the results of these studies, which shows the nominal detector efficiency and the efficiency loss from the dead time due to crosstalk, for each readout partition ($\eta$ number) in the detector.

\section{MITIGATION STRATEGIES FOR REDUCING CROSSTALK}
\begin{figure}[!ht]
\centering
\includegraphics[width=\columnwidth]{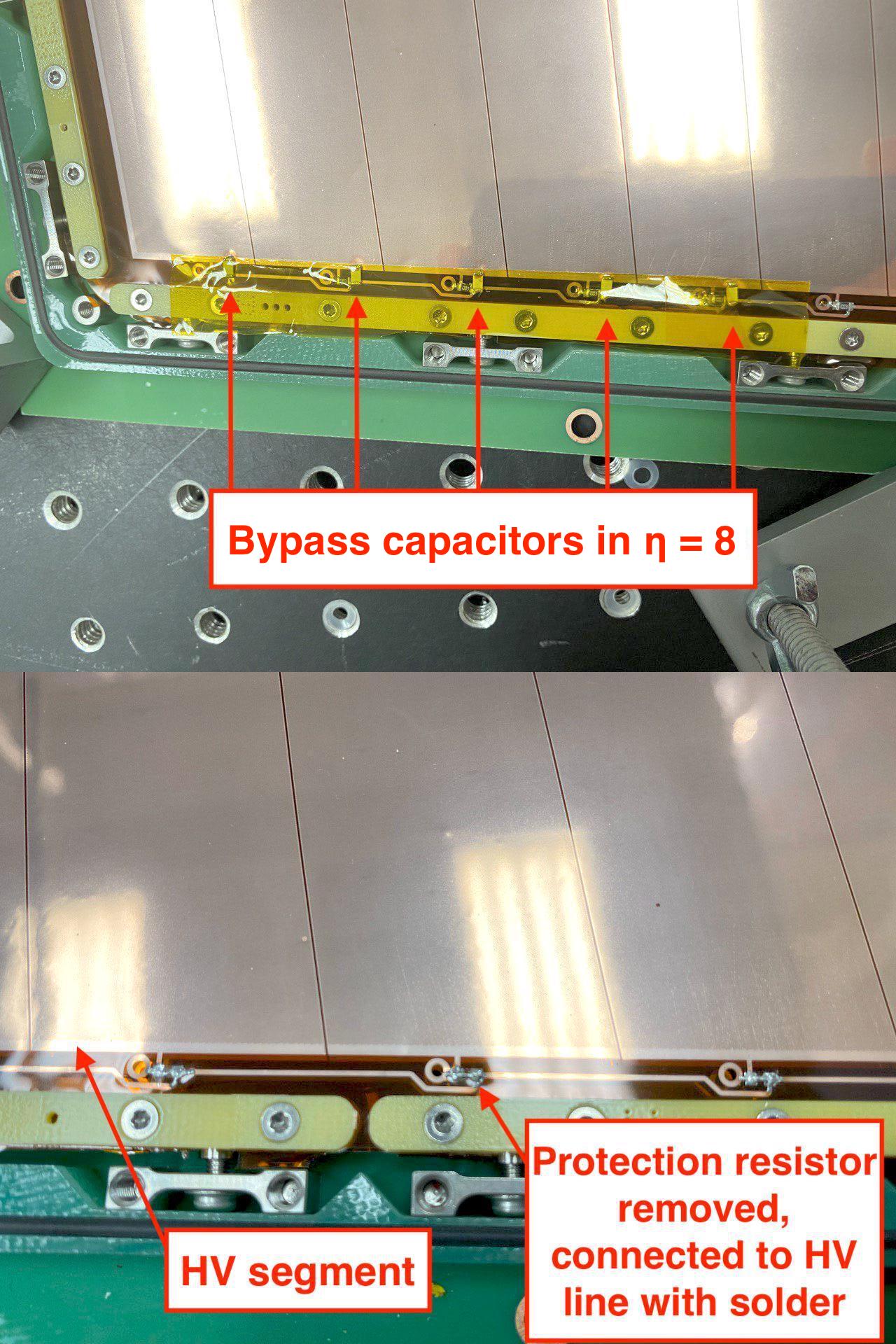}
\caption{\label{fig:GEM3Bmod}The bypass capacitors (covered with a layer of Kapton tape) installed in $\eta=8$ on GEM3B (top) and the HV segments connected in parallel with solder in $\eta=5$ on GEM3B (bottom).}
\end{figure}

Several mitigation strategies were tested to ameliorate the crosstalk: increasing the area of the HV segments on the bottom of the third GEM foil (GEM3B), both with and without a low-pass filter, and installing bypass capacitors in one $\eta$ segment. The idea is to reduce the impedance to ground for AC signals by increasing the capacitance between strips and the bottom of the GEM3 foil or by creating a direct AC path with a capacitor \cite{Hohlmann}. For the first study, we soldered five, 330 pF bypass capacitors in parallel with the protection resistors to the HV segments in the $\eta=8$ sector on the bottom of the third GEM foil, and removed the protection resistors in $\eta=5$, connecting these HV segments in parallel to increase the capacitance of the third GEM foil (see Fig.~\ref{fig:GEM3Bmod}).

\begin{figure*}[!ht]
\centering
\includegraphics[width=0.8\linewidth]{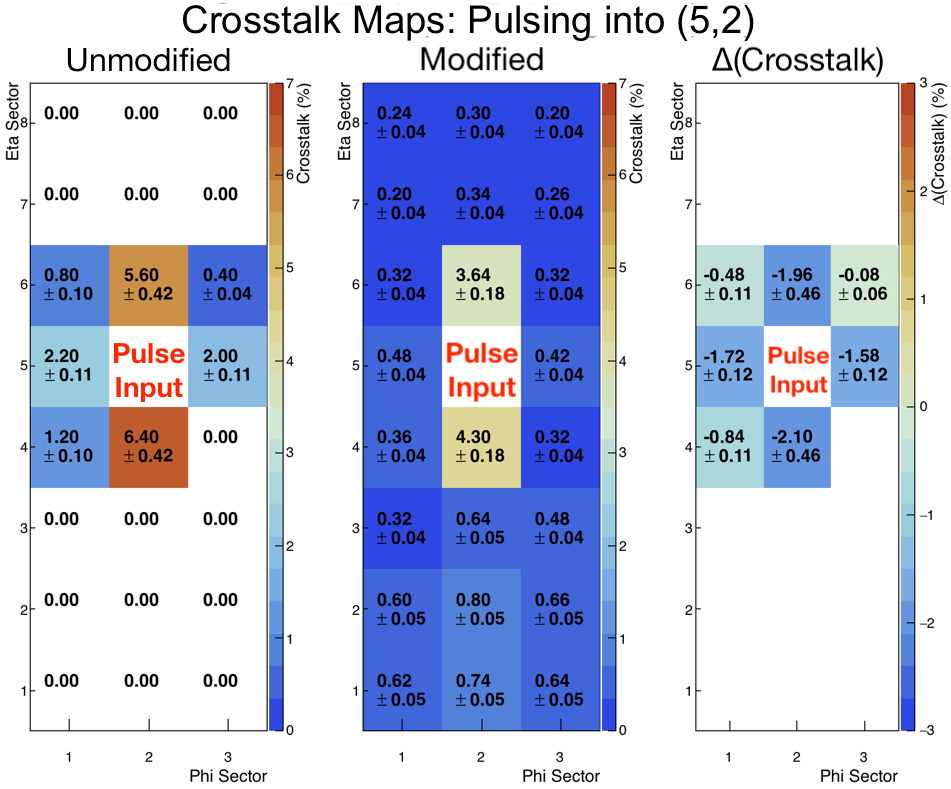}
\caption{\label{fig:summaryPlotAll52}Crosstalk maps of the observed crosstalk percentage for the unmodified, segmented GEM3B (left), the modified version of GEM3B where all HV segments are connected in parallel with solder, with HV filter and HV divider connected (center), and the change in the crosstalk percentage between the unmodified and the modified chamber (right), for pulse input into sector (5,2).}
\end{figure*}

Crosstalk maps were then taken for pulse inputs into each RO connector in $\eta = 5, 8$. We then repeated crosstalk measurements with all of the 37 HV segments on GEM3B connected in parallel with solder (in effect recreating an unsegmented GEM3B electrode). This configuration was measured both with a low-pass circuit and both with and without the HV divider. An example map for the configuration of all HV segments connected, a low pass circuit, and the HV divider, along with the original map in the baseline configuration, as well as a map that displays the total change in percentage, is presented in Fig.~\ref{fig:summaryPlotAll52}. We see that although new sectors farther away from the pulse injection are experiencing a small amount of crosstalk, there is an overall decrease in the sectors previously suffering from crosstalk. The results, quoted as a change in percentage of crosstalk amplitudes, are listed in Table \ref{tab:results}. 

\begin{figure*}[!ht]
\centering
\includegraphics[width=0.8\linewidth]{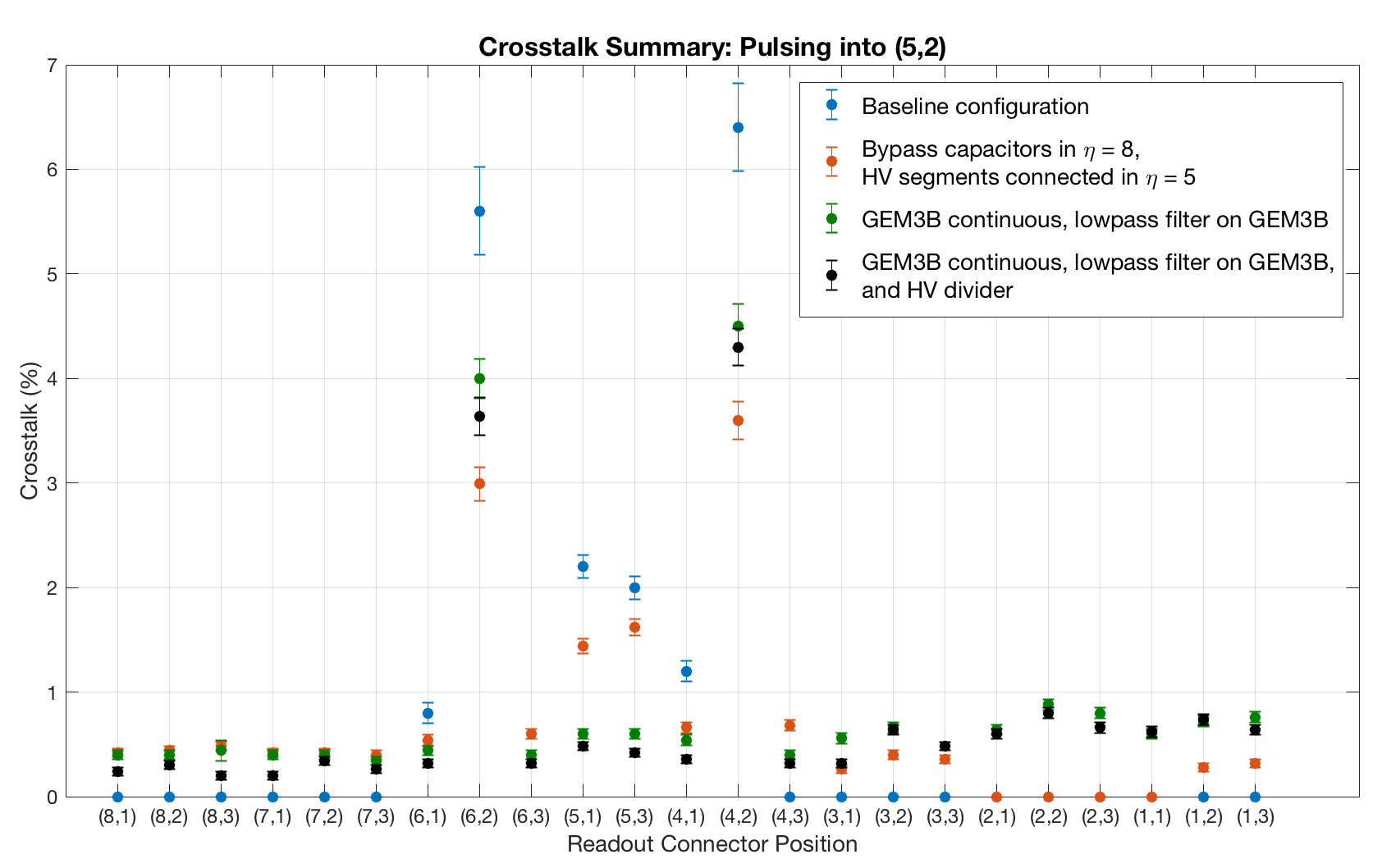}
\caption{\label{fig:summaryPlot}Summary plot of the observed crosstalk percentage in all sectors for pulse input into (5,2) for different mitigation measures.}
\end{figure*}

For a summary of all mitigation strategies for pulse input into readout sector (5,2), see Fig.~\ref{fig:summaryPlot}. It should be noted that a small value of crosstalk $\big(\hspace{-0.2cm}\lessapprox \hspace{-0.1cm}0.4\%\big)$ was observed in all $\phi$ partitions in all $\eta$ sectors after these modifications were made. This is expected because the contiguous GEM3 bottom can couple a small amount of crosstalk into all RO sectors. Overall, the average observed crosstalk is reduced by each mitigation strategy, with the largest decrease in crosstalk occurring when the third GEM foil is contiguous (i.e., protection resistors on the top-side of the foil, only), with the HV divider and low-pass filter connected.



\begin{table*}[!t]
\centering
\caption{\label{tab:results}Average Change in Crosstalk for All Sectors Previously Experiencing Crosstalk With Bypass Capacitors ($\eta=8$) and HV Segments Connected ($\eta=5$), and With All HV Segments Connected on GEM3B with HV Filter, Both With and Without an HV Divider}
\begin{tabular}{c| c c c} 
\hline
Pulsing into & Bypass Cap. \& HV segments	 & GEM3B Continuous, HV Filter 	& GEM3B Continuous, HV Filter \\
 & connected in $\eta=5$					 & (w/o HV Divider) 			& (w/ HV Divider) \\ [0.5ex]
\hline
$\eta =8$ & (-0.47$\pm$0.04)\%	& (-0.50$\pm$0.04)\%   	& (-0.53$\pm$0.03)\%\\
$\eta =5$ & (+0.03$\pm$0.04)\%	& (-0.05$\pm$0.05)\%   	& (-0.36$\pm$0.07)\%\\
$\eta =1$ & N/A					& (-0.17$\pm$0.04)\% 	& (-0.39$\pm$0.05)\%\\
\hline 	
Grand Average & (-0.22$\pm$0.03)\%& (-0.24$\pm$0.03)\% 		& (-0.43$\pm$0.03)\%\\
\hline
\end{tabular}
\end{table*}
\section{Summary and Conclusion}
During the initial performance testing of a prototype ME0, a triple-GEM detector proposed for the\\ Phase-2 Muon Upgrade of the CMS experiment, substantial crosstalk was observed in neighboring readout sectors. This motivated our investigations to better understand and mitigate this crosstalk, as outlined in this paper. By applying square voltage pulses to all 128 readout strips in a RO sector, and reading the signal out of each remaining RO sector, we observe that, in the version of the ME0 triple-GEM detector where GEM foils have HV protection resistors on both sides, a range of crosstalk between 0.24\%--6.40\% is seen across all $\phi$ partitions of a readout sector that is being pulsed, with crosstalk extending to the nearest neighboring $\eta$ segments. This crosstalk is due to the capacitive coupling between RO sectors and the coupling between RO sectors and the electrode at the bottom of GEM3. We see that while a small value of crosstalk is introduced into other RO sectors by making the bottom of the third GEM foil contiguous again, the crosstalk is successfully reduced, with a maximum average reduction from\hfill\\%
$(1.66\pm0.03)\%$ to $(1.11\pm0.02)\%$ with a low-pass filter and HV divider. Simulations of the frontend hybrid card ASIC circuit response indicate that without these modifications, we could expect a maximum efficiency loss of $\sim$6\% when operating the detector in CMS. Consequently, a design with a contiguous bottom electrode on GEM3 but double-segmented GEM1 and GEM2 foils has been adopted by the CMS muon group as the final design for mass production of GE2/1 foils (see Fig.~\ref{fig:quadrant} for the location of the GE2/1 chambers in the endcap of the CMS experiment), which face the same crosstalk issues as the ME0 module discussed here.

We note that the purpose of the initial double-segmentation in all three GEM foils was to limit discharge propagation and rate in the GE2/1 detector. Tests show that the discharge probability is reduced by three orders of magnitude \cite{Jeremie}. The effect on the discharge probability with the adopted configuration (with the third GEM foil segmented only on one side and the other two foils segmented on both sides) is currently under study. Preliminary results indicate that this ``mixed design" can simultaneously reduce discharge propagation and crosstalk with final results to be published soon.

\section*{Acknowledgments}
We gratefully acknowledge support from FRS-FNRS (Belgium), FWO-Flanders (Belgium), BSF- MES (Bulgaria), MOST and NSFC (China), BMBF (Germany), CSIR (India), DAE (India), DST (India), UGC (India), INFN (Italy), NRF (Korea), QNRF (Qatar), DOE (U.S.A.), and NSF (U.S.A.). We would also like to thank all members of the CMS GEM group for their contributions to this project.



%




\end{document}